\documentclass[aps,prl,twocolumn,superscriptaddress]{revtex4-2}

\usepackage{amsmath,amssymb,mathtools}
\usepackage{mathrsfs}
\usepackage{bbold}
\usepackage{epsfig}
\usepackage{graphicx,color}
\usepackage[colorlinks=true,linkcolor=blue,citecolor=blue]{hyperref}

\begin{document}

\title{Pairwise mode-locking in dynamically-coupled parametric oscillators}
\author{Leon Bello}
\affiliation{Department of Physics and BINA Center of Nanotechnology, Bar-Ilan University, 52900 Ramat-Gan, Israel}
\author{Marcello Calvanese Strinati}
\affiliation{Department of Physics, Bar-Ilan University, 52900 Ramat-Gan, Israel}
\author{Shai Ben-Ami}
\affiliation{Department of Physics and BINA Center of Nanotechnology, Bar-Ilan University, 52900 Ramat-Gan, Israel}
\author{Avi Pe'er}
\affiliation{Department of Physics and BINA Center of Nanotechnology, Bar-Ilan University, 52900 Ramat-Gan, Israel}

\date{\today}

\begin{abstract}
\textcolor{black}{Mode locking in lasers is a collective effect, where due to a weak coupling a large number of frequency modes lock their phases to oscillate in unison, forming an ultrashort pulse in time. We demonstrate an analogous collective effect in coupled parametric oscillators, which we term ``pairwise mode-locking'', where many pairs of modes with twin frequencies (symmetric around the center carrier) oscillate simultaneously with a locked phase-sum, while the phases of individual modes remain undefined. Thus, despite being broadband and multimode, the emission is not pulsed and lacks first-order coherence, while possessing a very high degree of second-order coherence. Our configuration is comprised of two coupled parametric oscillators within identical multimode cavities, where the coupling between the oscillators is modulated in time at the repetition rate of the cavity modes, with some analogy to active mode-locking in lasers. We demonstrate pairwise mode-locking in a radio-frequency (RF) experiment, covering over an octave of bandwidth with approximately 20 resonant mode-locked pairs, filling most of the available bandwidth between DC and the pump frequency. We accompany our experiment with an analytic model that accounts for the properties of the coupled parametric oscillators near threshold.}
\end{abstract}
\maketitle

The parametric oscillator (PO) is a central device in modern quantum optics - a fundamental type of oscillator whose internal parameters are modulated by an external drive, leading to parametric amplification~\cite{manley_rowe_1,rowe_2,pumping_tea,strogatz2007nonlinear}. Below the oscillation threshold, parametric oscillators are extensively used as sources of squeezed non-classical light, where the quantum fluctuations of one quadrature of the field are reduced below the vacuum (shot-noise) level, at the expense of increased fluctuations in the orthogonal quadrature~\cite{Lvovsky_2015, PhysRevA.29.408,JOSAB.4.001465,PhysRevA.30.1386}, with applications in metrology~\cite{PhysRevD.23.1693,Harry2010,Aasi2013,qdm}, basic quantum information~\cite{Ralph1999,Furusawa1998,Braunstein2000,s41377-018-0011-3} and quantum communication~\cite{PhysRevA.61.010303,s41467-018-03083-5}. Parametric oscillators have been proposed as scalable sources for continuous-variable (CV) cluster states~\cite{PhysRevLett.112.120505} for CV one-way quantum computation~\cite{PhysRevLett.101.130501}. Configurations of coupled parametric oscillators were explored in quantum information~\cite{furusawa, pfister, Howell2004}, quantum computing~\cite{PhysRevLett.112.120505} and coherent computing \cite{yamamoto, Inagaki2016b, s41534-017-0048-9, McMahon2016}.

In classical nonlinear optics, a parametric amplifier converts a pump field at frequency $\omega_p$ into a pair of signal and idler fields ($s$ and $i$) such that $\omega_s + \omega_i = \omega_p$.
\textcolor{black}{Below threshold}, when the process \textcolor{black}{is spontaneously generated by single pairs of photons}, each field appears as thermal \textcolor{black}{noise} \cite{PhysRevA.36.3464}, but the radiation produced is two-photon coherent \cite{PhysRevLett.93.023005, PhysRevLett.94.073601, PhysRevLett.94.043602}, i.e., the spectral phase of each frequency mode is random, but the sum of phases $\phi_s + \phi_i = \phi_p$ of all signal-idler pairs is well defined and highly coherent \cite{Villar2005, Coelho2009}. Normally, non-linear effects are weak, indicating that intense pump fields are required to obtain appreciable down-conversion powers.
As such, parametric oscillators, where the parametric amplifier is incorporated inside a high-finesse cavity \cite{PhysRevA.29.408}, are used to critically reduce the required pump power. \textcolor{black}{Due to the cavity, the emitted field is spectrally structured into a discrete comb of frequency modes that are separated by the cavity repetition rate, where below the oscillation threshold all available signal-idler pairs emit two-mode squeezed vacuum, which is therefore highly multimode.} Above the oscillation threshold mode competition drastically narrows the down-conversion bandwidth, ultimately to a \emph{single} signal-idler pair \cite{Fabre2000, YarivBook}.
\textcolor{black}{Mode competition also exists in laser, where it pushes the system towards single-mode operation, yet lasers can demonstrate extremely broadband oscillation via mode locking \cite{haus_modelocking}. In mode locking, weak coupling between the modes within the gain causes an effective phase transition \cite{Schwartz_2013, Weill_2007} from a single-mode oscillation to a macroscopic number of oscillating modes with a locked relative phase, collapsing the field in time to an ultrashort pulse. In a rough sense, it is most efficient for the laser to store energy within the medium, and then release it in a short intense pulse at the times when the losses are minimal. 
This type of mechanism does not exist in a parametric oscillator, since the nonlinear gain is instantaneous and lacks any ability to store energy. Thus, any pump energy that is not immediately converted to signal-idler pairs is lost. Thus, it is inefficient for a parametric oscillator (in the absence of additional non-linearities, e.g. Kerr non-linearity) to support a pulsed oscillation when the pump is continuous \cite{Yu2016}.}

\begin{figure}[t]
	\centering
	\includegraphics[width=0.35\textwidth]{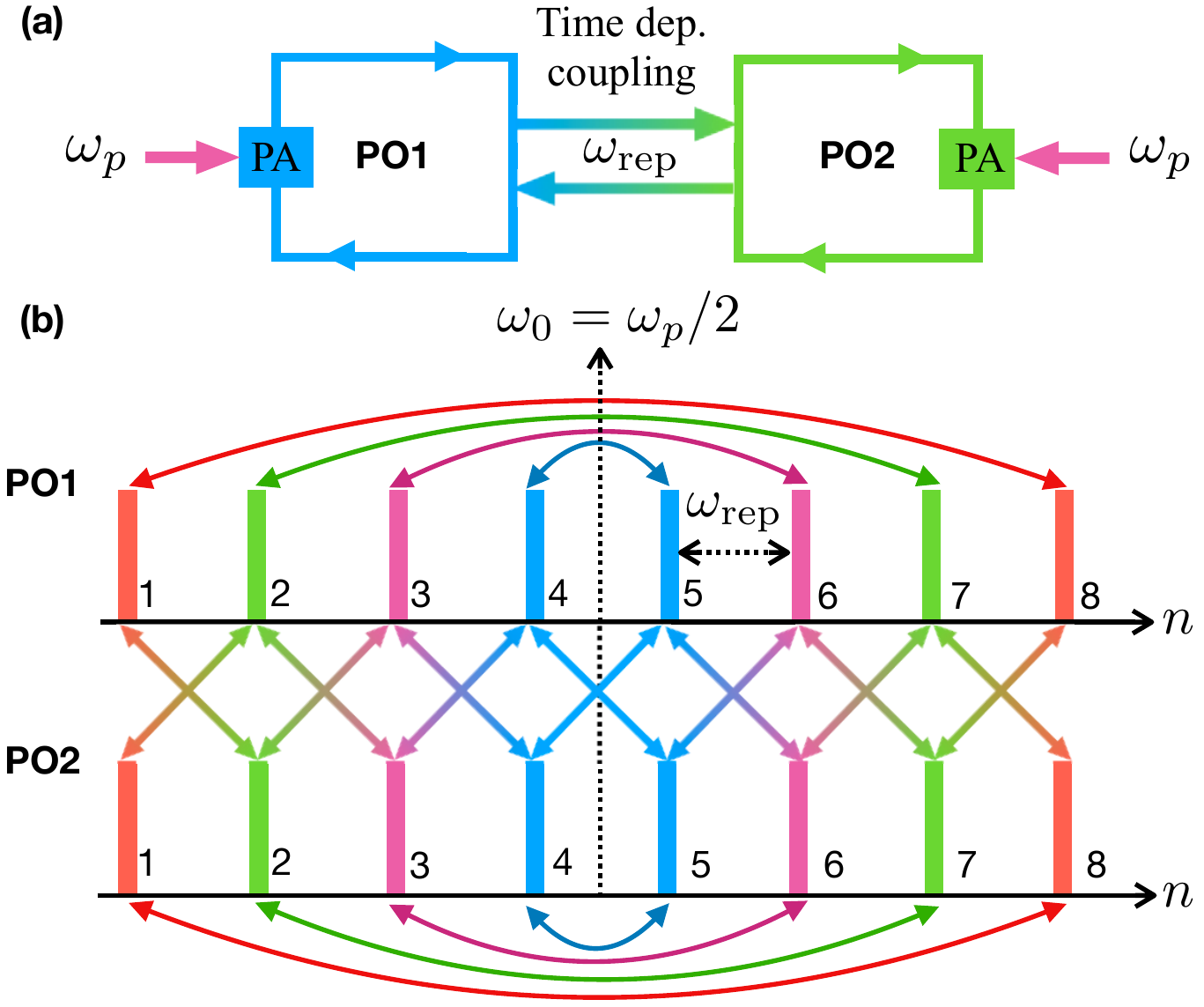}
	\caption{\textbf{(a)} Schematic of a two coupled doubly resonant cavities driven by a single-mode pump. \textbf{(b)} Illustration of the configuration of modes in our system. For illustration purposes, the two parametric oscillators (PO1 and PO2) have $8$ longitudinal modes each labeled by $n=1,\ldots,8$, equally spaced by $\omega_{\rm rep}$. Bars with the same color indicate the signal-idler pairs of each oscillator. Diagonal lines between the bars indicate coupling between the different cavities, whereas curved lines above and below the bars indicate signal-idler pairs. The vertical black dashed line marks the central frequency $\omega_0 = \omega_p/2$. In the experiment, the coupling additionally has a DC component coupling modes of the same frequency between the cavities (not drawn), which does not change the dynamics considered here.}
	\label{fig:mode_couplings}
\end{figure}

We present here a novel configuration of two multi-mode \emph{AC-coupled} parametric oscillators that produce a highly multi-mode coherent oscillation above threshold. \textcolor{black}{Previously we described the coherent dynamics of two coupled single-mode parametric oscillators, which leads to steady-state emission of persistent coherent beating \cite{PhysRevLett.123.083901, PhysRevA.100.023835}. Here we describe how modulation of the coupling causes a macroscopic effect of \emph{pairwise mode-locking - a collective multimode oscillation} across all the available modes of the two coupled cavities, similar to the generation of a a broadband optical frequency comb by mode-locking in lasers. Furthermore, the observed oscillation in this pairwise mode-locked regime is fundamentally different in both frequency and time from the simple coherent beating pattern in the single mode case}. The key concept in our configuration is the modulation of the coupling between two parametric oscillators in time at a frequency that is an integer multiple of the repetition rate $\omega_{\rm rep}$ of the cavities (Fig.~\ref{fig:mode_couplings}). This actively couples frequency modes of one oscillator to neighbouring modes \emph{of the other oscillator}, inducing a bright and broadband parametric oscillation. In spite of the similarity of this scheme to active mode-locking in lasers. \textcolor{black}{Since the oscillation is above threshold, each mode on its own is a coherent oscillation with a well defined phase and amplitude. However, the phases of different modes are uncorrelated, and thus the overall oscillation lacks first-order coherence,} but demonstrates a very high degree of second-order coherence, where all signal-idler pairs are complex conjugates of each other. Since our oscillation shows high \textcolor{black}{mutual} coherence as pairs of modes but not as single modes, hence we term our method ``pairwise'' mode-locking.

We demonstrate our configuration in a radio-frequency (RF) experiment, using simple off-the-shelf components.
The reason for the choice of an RF platform is that it provides a simple and nearly ideal platform to explore coherent phenomena and to directly observe the oscillation in both time and frequency. In addition, setups that are difficult to design and implement in optics can sometimes be trivially implemented in RF, with substantially less resources. \textcolor{black}{Although the RF experiment is purely classical, and cannot capture quantum effects, such as sub-vacuum squeezing and photon counting, it demonstrates well the coherent physics involved in the coupling above threshold.} 

\begin{figure}[t]
	\centering
	\includegraphics[width=0.4\textwidth]{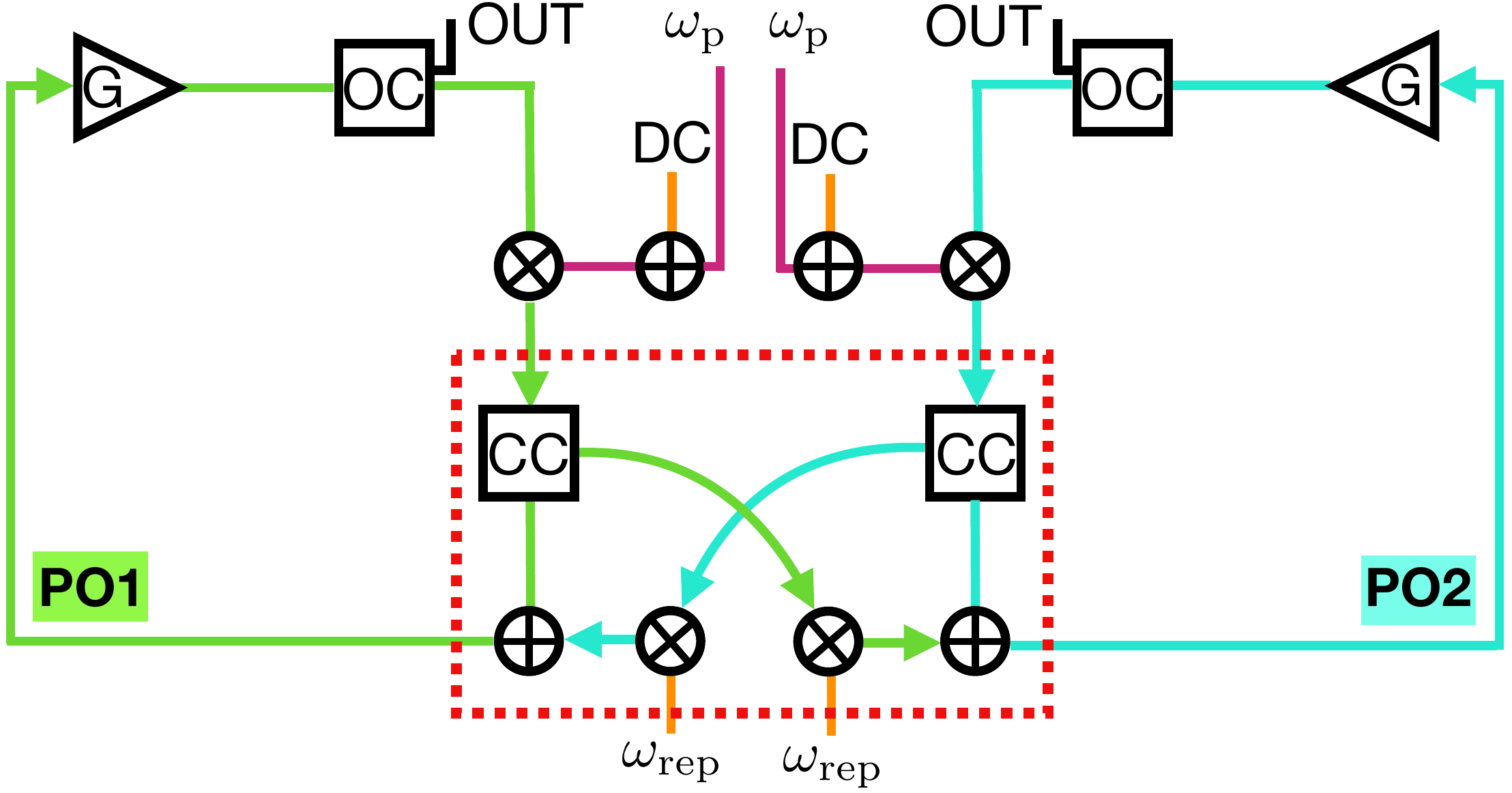}
	\caption{Each oscillator is comprised of the following components by Mini-Circuits: The parametric amplifier is realized by a RF frequency mixer (ZX05-10-S+) denoted by ``$\otimes$'', and driven at approximately $\omega_{\rm p}/2\pi=720\,{\rm MHz}$ by a RF synthesizer (Agilent N5181A) summed together with a DC offset on a bias-tee (ZFBT-4R2G-FT-+), denoted by ``$\oplus$''. Since the parametric gain of the mixer was insufficient to cross oscillation threshold, a broadband (linear) low-noise amplifier ``G'' (ZX60-P105LN) was added to mitigate some of the losses, but could not induce oscillations on its own. An output coupler (OC) (ZFDC-15-5) couples the oscillation out for observation. The time-dependent coupling at frequency $\omega_{\rm rep}$ is represented in the red-dashed box. \textcolor{black}{Another coupler (CC) is used to couple between the oscillators.} To control the amount of power coupled into the other cavity, we pass the signal through a voltage-controlled variable attenuator implemented using another mixer, and inject it into the other oscillator using a power combiner denoted by ``$\oplus$'' (ZAPD-2-252-S+).}
	\label{fig:scheme}
\end{figure} 

Our experimental setup is illustrated in Fig.~\ref{fig:scheme}. We implement the two multi-mode parametric oscillators by using two broadband RF cavities, coupled with a time-dependent coupling mechanism. The oscillators, labeled PO1 and PO2, are identical in components and demonstrate very close resonances. \textcolor{black}{In the experiment, remnant DC coupling is present and locks the resonances together when coupled}. We realize the oscillators with standard RF components, and $6$-meter-long coaxial cables in a ring configuration, forming two resonators with a repetition rate  $\omega_{\rm rep}/2\pi\simeq15\,{\rm MHz}$. Each oscillator is pumped by a single frequency pump at $\omega_{\rm p}/2\pi=720\,{\rm MHz}$ and the coupling between them is achieved using a power splitter, which injects a controlled amount of signal from each cavity into the other one. \textcolor{black}{The cavity determines a set of resonant modes, and by tuning the pump to a resonant mode, we parametrically ampilfy all mode-pairs that sum up to the pump-mode frequency. Changing the pump frequency allows us to control the bandwidth and the resonant modes to be amplified.}
The coupled signal is then modulated in time with a function generator at $\omega_{\rm rep}$, as indicated in Fig.~\ref{fig:scheme}. \textcolor{black}{The intensity and frequency of the modulation dictates the strength and the range of the coupling. In general the coupling could be modulated at any integer multiple of the repetition rate and not necessarily be monochromatic, and general coupling could be devised to create different coupling topologies \cite{Schwartz_2013, Onodera2020}.}

\begin{figure}
	\centering
	\includegraphics[width=0.45\textwidth]{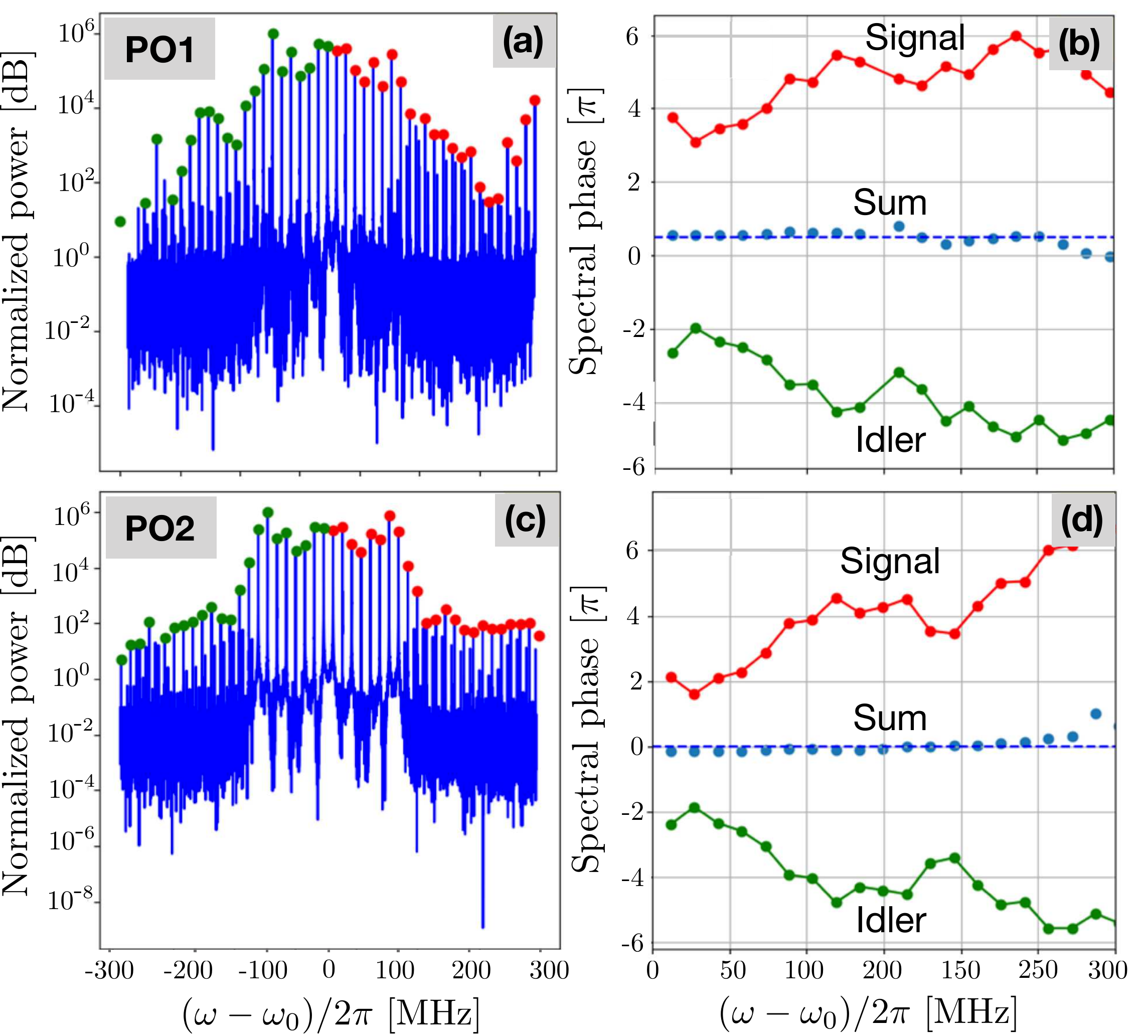}
	\caption{\textbf{(a),(c)} Power spectrum \textcolor{black}{normalized to the noise floor} of the broadband oscillation in PO1 and PO2, respectively. Signal modes are indicated with red dots, and idler modes with green dots. \textbf{(b),(d)} Spectral phases of the two oscillations in PO1 and PO2, respectively. Blue dots indicate the sum of the signal (red dots) and idler (green dots) phases, with the dotted blue line indicating the pump phase.}
	\label{fig:broadband_parametric_oscillation}
\end{figure}

\begin{figure}
	\centering
	\includegraphics[width=0.35\textwidth]{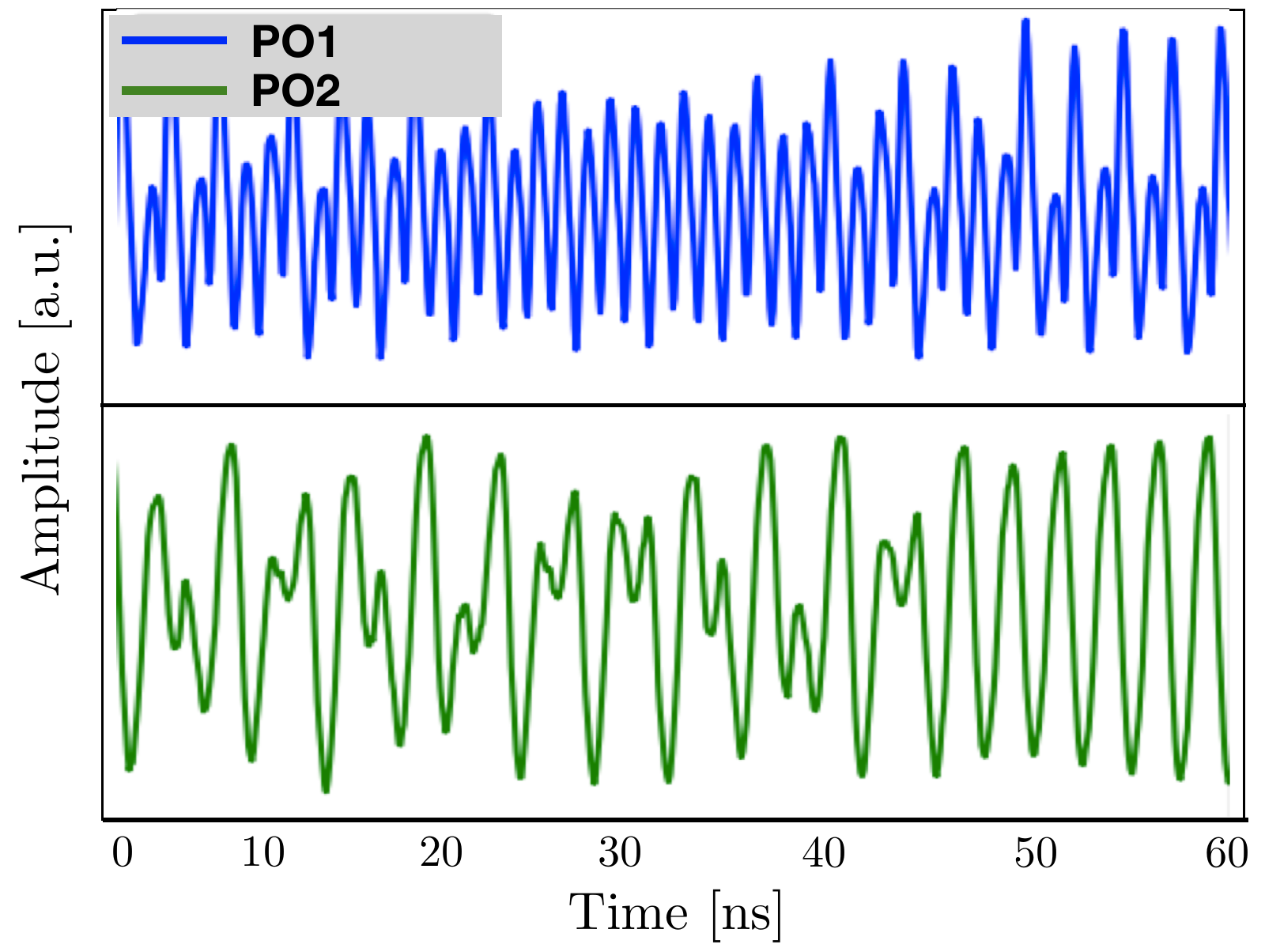}
	\caption{Oscillation in time for PO1 (blue data) and PO2 (green data), from which the spectral properties (Fig.~\ref{fig:broadband_parametric_oscillation}) are extracted. As evident, no pulsing is observed.}
	\label{fig:time_series}
\end{figure}

When the coupling modulation frequency is within few percent of the repetition rate $\omega_{\rm rep}$ the oscillation becomes very broadband, filling nearly the entire available bandwidth, as shown in Fig.~\ref{fig:broadband_parametric_oscillation}(a),(c). Evidently, the modes oscillate symmetrically around the carrier frequency $\omega_0=\omega_p/2$, in pairs with conjugate phases that sum up to the pump phase all across the oscillation spectrum, showing high pairwise second-order coherence. In time, this spectral symmetry indicates that the oscillation is on a single quadrature of the electric field, i.e. $E^{(j)}(t)=X^{(j)}(t)\cos(\omega_0 t)$ with no $Y^{(j)}\sin(\omega_0 t)$ quadrature component. Although the oscillation is very broadband, we measure no correlation between the spectral phases of different signal-idler pairs [Fig.~\ref{fig:broadband_parametric_oscillation}(b),(d)], which is a clear indication of the lack of first-order coherence \textcolor{black}{between different modes} and therefore \textcolor{black}{also the absence of pulsed oscillations in time}, as can be also directly observed on an oscilloscope (Fig.~\ref{fig:time_series}).
Our findings show that our technique is profoundly different from active mode-locking in lasers. The pairwise coherent oscillation, although broadband, is first-order incoherent and appears continuous in time, and thus efficiently utilizes the pump. 
Despite the conceptual difference, a useful and elegant analogy between pairwise and standard mode-locking exists. We model our system by two identical cavities, driven by a pump field that is resonant on the $N$-th mode of the oscillators, at frequency $\omega_p = N\,\omega_{\rm rep}$, where $N$ is a positive integer (Fig.~\ref{fig:mode_couplings}). For simplicity, we set $\omega_{\rm rep}=1$. The pump field is down-converted into a pair of signal and idler modes, at frequencies $\omega$ and $N - \omega$, respectively. The dynamics of the modes in the two cavities is captured by the complex slow-varying amplitudes~\cite{PhysRevA.100.023835} $A^{(1)}_\omega$ and $A^{(2)}_\omega$, at frequency $\omega$ in cavities PO1 and PO2, respectively. The energy of the pump is converted into conjugate signal-idler pairs, ~\cite{mandel_wolf_1995, caves_2-mode}: $A^{(j)}_\omega=(A^{(j)}_{N-\omega})^*$, for all $\omega$ and $j=1,2$. These relations reflect the fact that the sum of phases of each signal-idler pair sum up to the pump phase, taken to be zero. In addition, the lack of first-order coherence for different modes implies that they are uncorrelated $\langle A^{(j)}_\omega(A^{(j')}_{\omega'})^*\rangle=I^{(j)}_\omega\delta_{j,j'}\delta_{\omega,\omega'}$, where we define the steady-state power spectrum $I^{(j)}_\omega \equiv \langle|A^{(j)}_\omega|^2\rangle$ [Fig.~\ref{fig:broadband_parametric_oscillation}(b),(c)], where the expectation value is an ensemble average over all possible spectral phases. The modes in the two cavities are coupled by a time-dependent coupling, modulated at $m\omega_{\rm rep}$, with $m$ integer. Here we focus on $m = 1$ (nearest-neighbor coupling).

The dynamics in PO1 and PO2 are described by $2N$ coupled first-order ordinary differential equations:
\begin{multline}
	\cfrac{d}{dt}A^{(j)}_\omega = \left(G_\omega - \beta \sum_{\omega'} A_{N-\omega'}^{(j)} A_{\omega'}^{(j)}\right) \left(A^{(j)}_{N - \omega}\right)^* \\ +\cfrac{\delta}{2}\left[A^{(k)}_{\omega + m}+A^{(k)}_{\omega-m}\right] ,
\label{eq:quadrature_equations_of_motion_1}
\end{multline}
for $j\neq k=1,2$, where $m < \omega \le N - m$ with $N$ odd, and $G_\omega=h/8-g_\omega/2$ is the net gain per round-trip of the $\omega$-mode, dictated by the parametric gain $h$ and the loss term $g_\omega$. Gain saturation due to the pump depletion is denoted by $\beta$. The strength of the coupling between mode $\omega$ in PO1 and modes $\omega \pm m$ in PO2 is denoted by $\delta$ .
The first terms in the right-hand side of Eq.~\eqref{eq:quadrature_equations_of_motion_1} describe the parametric amplification for each individual signal-idler pair at frequency $\omega$ and $N-\omega$. In the uncoupled case ($\delta=0$), these independent pairs compete for the gain resources and only the one associated with the largest net gain $G_\omega$ oscillates. However, when $\delta\neq0$, the coupling connects all pairs at different frequencies. Consequentially, the most efficient mode is now a \textcolor{black}{broadband} combination of all signal-idler pairs, whose spectrum approximately mimics the spectral loss function. To show this theoretically, we derive from Eq. \eqref{eq:quadrature_equations_of_motion_1} an equation for the spectrum $I^{(j)}_\omega$ of the oscillation near threshold, where gain saturation is negligible~\footnote{See supplemental material.}
\begin{equation}
\frac{\delta^2\omega_{\rm rep}^2}{2}\frac{d^2}{d\omega^2}I_\omega^{(j)}+(4G_{\omega}^2+2\delta^2)I_\omega^{(j)} = 0 ,
\label{eq:equationformodelocking7}
\end{equation}
in a direct analogy to the spectral amplitude of pulses in active mode-locking of lasers~\cite{haus_modelocking}. \textcolor{black}{To solve Eq.~\eqref{eq:equationformodelocking7}, we assume} a simple parabolic dependence of the spectral net gain $G_\omega^2 \approx G_0^2 - \omega^2/2\sigma^2 $, where $\sigma$ is the gain bandwidth, \textcolor{black}{which} yields from Eq.~\eqref{eq:equationformodelocking7} a Gaussian spectrum $I^{(j)}_\omega \sim e^{-\omega^2/\Delta^2}$ with spectral width $\Delta^2 = \delta \omega_{\rm rep} \sigma$ and steady-state gain of $G_0^2 = \delta \omega_{\rm rep}/4\sigma-\delta^2/2$, similarly to active mode-locking in lasers. \textcolor{black}{The parabolic spectral gain was assumed here for simplicity of the analytic solution. Other gain profiles are possible as well, leading to more general spectral forms of the pulses, but keeping the bandwidth discussed here.}

\begin{figure}
	\centering
	\includegraphics[width=0.48\textwidth]{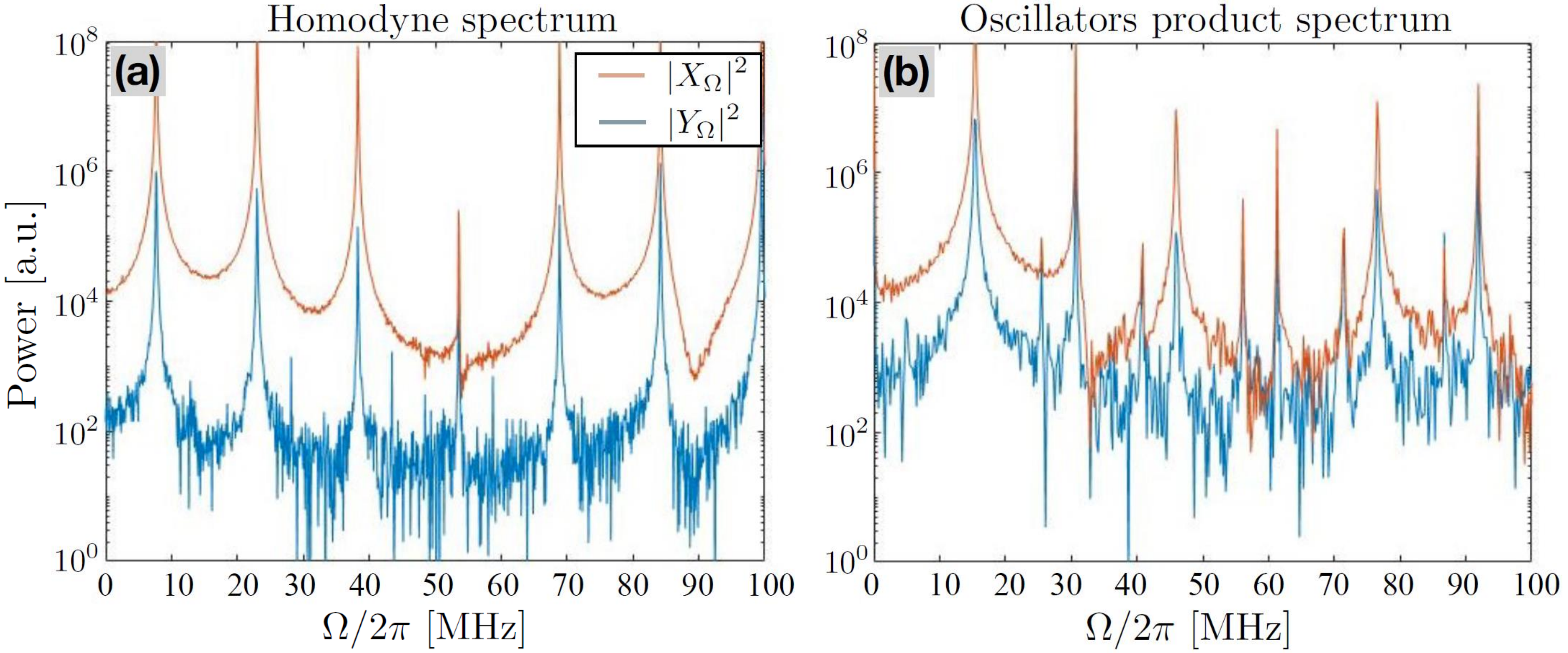}
	\caption{\textbf{(a)} The \textcolor{black}{amplified} ($|X_\Omega|^2$, orange) and attenuated ($|Y_\Omega|^2$, blue) quadrature spectra of the homodyne signal obtained by multiplying the oscillator signal with a phase-coherent LO at half the pump frequency. \textbf{(b)} Spectra of the product signal obtained by multiplying the oscillators together. }
	\label{fig:homodyne_spectrum}
\end{figure}

Let us consider the implications of the fact that the oscillation is broadband and on a single quadrature \textcolor{black}{due to the phase-dependent amplification}. Normally, the quadrature content of a signal is evaluated by homodyning against a local oscillator (LO) at the center carrier frequency $\omega_0$, acting as a quadrature reference, so for example $X^{(j)}(t) = \left \langle E^{(j)}(t) \cos(\omega_0 t) \right \rangle $, where the angle brackets denote averaging over fast-varying terms. Here, since the signals $E^{(j)}(t)$ are broadband, so is the homodyne result. \textcolor{black}{In frequency domain, the quadrature content is easily seen as the interference of pairs of modes $X^{(j)}_\Omega = \frac{1}{2}(A^{(j)}_{N/2 + \Omega} + A^{*(j)}_{N/2 - \Omega})$ and $Y^{(j)}_\Omega = \frac{1}{2i}(A^{(j)}_{N/2 - \Omega} - A^{*(j)}_{N/2 - \Omega})$}. In Fig.~\ref{fig:homodyne_spectrum}(a) we show the spectrum of the homodyne output for two LO phases - the \textcolor{black}{amplified} ($|X_\Omega|^2$) and the \textcolor{black}{attenuated} ($|Y_\Omega|^2$) quadratures, with $>20$dB difference, \textcolor{black}{a clear indication of the phase-dependent amplification.} \textcolor{black}{Each peak in the homodyne spectrum corresponds to a different frequency pair, which interfere constructively or destructively depending on the local oscillator phase. The strong phase-dependence of the spectrum is a direct result of the strong second-coherence of the signal, i.e.  the sum of phases of each signal-idler pair is highly correlated and equal to half the pump phase.}

The quadrature reference does not need to be narrowband \cite{Patera2009}. Since both signals of PO1 and PO2 are of a single quadrature, they could serve as local oscillators for one another $X_{\tau}(t) = \left \langle E^{(1)}(t)E^{(2)}(t+\tau) \right \rangle$, where the delay between them determines the measured quadrature. \textcolor{black}{Fig.~\ref{fig:homodyne_spectrum}(b) shows the spectrum of the cross-correlation between the two oscillators, for two different delays, measuring the \textcolor{black}{amplified} ($|X_\Omega|^2$) and attenuated ($|Y_\Omega|^2$) quadrature spectra}. Mixing the two outputs also shows a clear difference of $>15$dB between the two quadrature phases, but its structure is inherently different from the standard homodyne with a single frequency LO. 
Each \textcolor{black}{spectral component is due to the mixing between many different frequency modes, all with the same spectral separation. For example, the peak at DC is due to the mixing between all same frequency modes, the peak at $\omega_{\rm rep}$ is due to the mixing between modes one repetition rate apart, and so on.} Every frequency within this cross-homodyne is a collective result of mixing the entire combs with the relevant frequency offset. We should emphasize that although our results demonstrate the difference in power between the two quadratures, they do not demonstrate squeezing, \textcolor{black}{but only the phase-dependent nature of the parametric amplifier}. \textcolor{black}{Characterization of the squeezing requires comparing the attenuated quadrature noise to the device noise floor, which is not demonstrated}.

In summary, we presented a new kind of broadband parametric source, comprising of a pair of parametric oscillators, coupled with a time-modulated coupling. We experimentally demonstrated that the generated oscillation is very broadband, yet lacks first-order coherence - the different frequency pairs are distinct, and their phases are unconstrained and uncorrelated. This source of bright, broadband parametric radiation can be a key enabler for a variety of applications, e.g. quantum information~\cite{furusawa,pfister}, communication protocols~\cite{peer_ocdma,PhysRevA.61.010303,Ralph1999,marandi2012all}, bright quantum frequency combs \cite{Valcarcel2006, Pinel2012, Gerke2015}, CV cluster states sources~\cite{pfister, Wang2014, Yukawa2008, Yokoyama2013}, noise-radar schemes \cite{Axelsson2004, Lukin2001, Tan2008} and quantum metrology~\cite{PhysRevD.23.1693,Ruo_Berchera_2015,qdm,Aasi2013,Harry2010, Treps2002}.
From a more fundamental point of view, this is a source of radiation that is conceptually different from lasers~\cite{siegman_modes} and standard parametric oscillators~\cite{yariv1991optical}. \textcolor{black}{On the one hand, unlike a mode-locked laser, the different modes are not correlated in phase. On the other hand, unlike squeezed vacuum, the oscillation is not incoherent noise, but rather a combination of many coherent, though uncorrelated, frequency pairs.}

\begin{acknowledgments}
\vspace{0.2cm}
\textit{Acknowledgments}.
We are grateful to our lab members A. Kahana, M. Meller, I. Parshani, and Y. Michael for fruitful discussions. A. P. acknowledges support from the Israel Science Foundation (ISF) Grants No. 44/14 and U.S.-Israel Binational Science Foundation (BSF) Grant No. 2017743.  M.~C.~S. acknowledges support from the ISF Grants No.~231/14, 1452/14, and~993/19, and BSF Grants No.~2016130 and~2018726.
\end{acknowledgments}


%

\end{document}